\documentclass[aip,jap,groupedaddress,reprint]{revtex4-1}

\usepackage{graphicx}
\usepackage{amsmath}
\usepackage[colorlinks]{hyperref}
\hypersetup{
   pdftitle={Storage and on-demand release of microwaves using superconducting resonators with tunable coupling},
   pdfauthor={Mathieu Pierre, Ida-Maria Svensson, Sankar Raman Sathyamoorthy, G\"{o}ran Johansson, and Per Delsing},
   pdfkeywords={superconductivity, microwave circuits, superconducting resonator, tunable coupling},
   urlcolor=black,
   citecolor=blue,
   linkcolor=blue
   }

\begin{document}
\title{Storage and on-demand release of microwaves using superconducting resonators with tunable coupling}
\author{Mathieu \surname{Pierre}}
\email[]{mathieu.pierre@lncmi.cnrs.fr}
\author{Ida-Maria \surname{Svensson}}
\author{Sankar Raman \surname{Sathyamoorthy}}
\author{G\"{o}ran \surname{Johansson}}
\author{Per \surname{Delsing}}
\email[]{per.delsing@chalmers.se}
\affiliation{Department of Microtechnology and Nanoscience (MC2), Chalmers University of Technology, SE-412\,96 G\"{o}teborg, Sweden}

\begin{abstract}
We present a system which allows to tune the coupling between a superconducting resonator and a transmission line.
This storage resonator is addressed through a second, coupling resonator, which is frequency-tunable and controlled by a magnetic flux applied to a superconducting quantum interference device (SQUID).
We experimentally demonstrate that the lifetime of the storage resonator can be tuned by more than three orders of magnitude. A  field can be stored for 18\,$\mu$s when the coupling resonator is tuned off resonance and it can be released in 14\,ns when the coupling resonator is tuned on resonance. 
The device allows capture, storage, and on-demand release of microwaves at a tunable rate.
\end{abstract}

\maketitle

At the interface between quantum optics and integrated electronics, superconducting circuits constitute a flexible and scalable platform for quantum information processing based on the manipulation of qubits and microwave photons \cite{wallraff2004,hofheinz2009,eichler2011}.
One of the main challenges is to develop basic tools required to manipulate photons at the single photon level. Such functionalities include, for instance, photon generation \cite{houck2007}, detection, routing \cite{hoi2011}, and storage \cite{yin2013}.

A scalable architecture of a circuit quantum electrodynamics (circuit-QED) experiment can have the structure of a quantum network \cite{kimble2008}. In this scheme propagating photons carrying bits of quantum information travel between nodes where quantum information is processed through the interaction of these photons with various quantum systems. Such quantum systems can be superconducting quantum bits, spin ensembles, quantum dots, or mechanical oscillators. Despite recent progress in achieving a strong coupling with propagating photons, it is convenient to embed the quantum system in a cavity where its coupling to the field is resonantly enhanced \cite{wallraff2004}. In this scheme, the cavity must be able to exchange photons with its surrounding, for example, a microwave transmission line. 
However, the ability for a cavity to store photons for a long time is not compatible with its ability to release them fast, since these two time scales are linked.

In this article, we present a superconducting resonator which features a controllable coupling to a transmission line where photons can be emitted from the cavity, or inversely where incident photons can be fed to the cavity.  
The coupling of the resonator to its environment can therefore be dynamically tailored to a specific goal, either preserving the coherence of the quantum system or, on the contrary, enabling its fast measurement and control.
Adding new features to a basic component of circuit-QED experiments such as a superconducting resonator gives rise to new possibilities. For example, making a resonator tunable in frequency \cite{wallquist2006,sandberg2008,palacios2008} has led to parametric amplifiers \cite{yamamoto2008,sundqvist2013} and oscillators \cite{wilson2010,wustmann2013}.

\begin{figure}[b]
\begin{center}
\includegraphics[width=1\columnwidth]{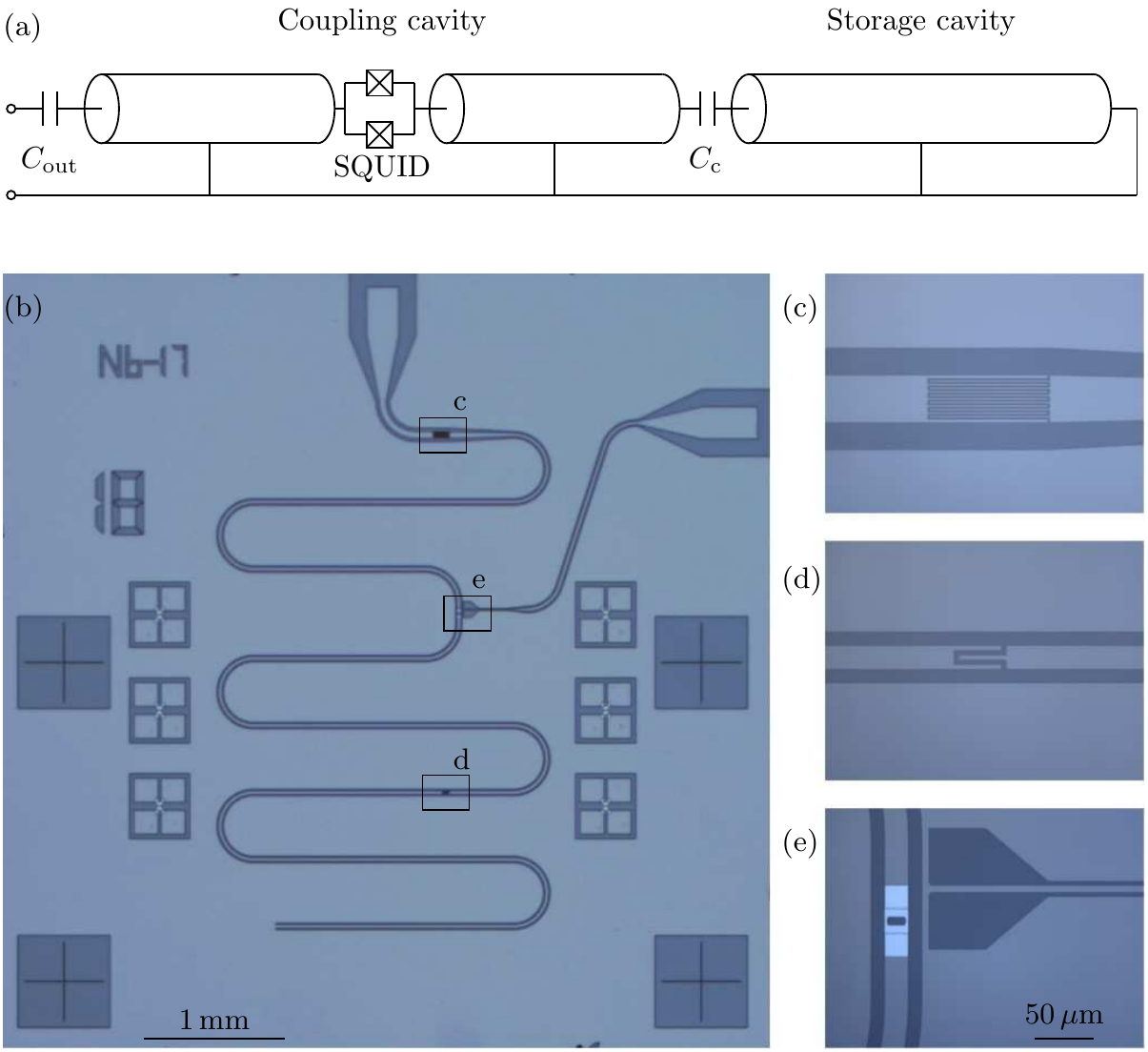}
\caption{a) Schematic of the device. A $\lambda/4$ transmission line resonator (storage cavity) is connected to the input port via a $\lambda/2$ resonator with a SQUID in its center (coupling cavity). b--e) Optical microscope images of sample A. Subfigures c, d and e have the same scale. b) Image of the $5\times5\,\mathrm{mm}^2$ chip, showing the coplanar waveguide resonators, from the input port on top to the grounded end of the storage cavity. The circuit is etched in a thin film of niobium. c) Capacitor between the input transmission line and the coupling cavity. d) Capacitor between the two resonators. e) Aluminum SQUID in a gap of the niobium center conductor. The line on the right side is used to induce magnetic flux in the SQUID loop.} 
\label{fig1}
\end{center}
\end{figure}

Using a tunable coupling for a resonator effectively tunes its external Q-value, and it is a way to control the emission and engineer the shape \cite{pechal2013,wenner2013} of photons initially stored in a cavity. This is useful to transfer a quantum state with arbitrarily high fidelity between distant cavities \cite{korotkov2011}. 
Moreover, a tunable coupling can also be used for quantum bath engineering. It indeed enables the control of the damping of a quantum system embedded in a cavity. This can be used for the fast initialization of a quantum bit \cite{jones2013} or for the creation of arbitrary strongly squeezed states of the field \cite{didier2013}.

Our scheme uses a $\lambda/2$ frequency-tunable resonator \cite{wallquist2006,sandberg2008,palacios2008} intercalated between a $\lambda/4$ fixed-frequency resonator and a transmission line (Fig.~\ref{fig1}). 
The fixed resonator will be referred to as the storage cavity, whereas the frequency-tunable resonator will be referred to as the coupling cavity.
The coupling of the storage cavity to the transmission line depends on the transmission of microwaves through the coupling cavity.
The latter behaves as a Fabry-P\'{e}rot cavity and thus its transmission is frequency-dependent.
The key feature of the system is that the resonance frequency of the coupling cavity can be adjusted with respect to the resonance frequency of the storage cavity, which leads to a variable coupling between the storage cavity and the transmission line. It is the detuning between the two cavities which determines the strength of the coupling.

The tunability in frequency of the coupling cavity is obtained by introducing a superconducting quantum interference device (SQUID) at its voltage node, \textit{i.e.} in its middle. 
The tunable inductance of the SQUID provides a variable contribution to the resonator total inductance, and thus makes its resonance frequency tunable. Moreover, its inductance can be tuned extremely fast.

The coplanar transmission line resonators are made to have a characteristic impedance $Z_0=50\,\Omega$. Their center conductor as well as the ground planes are made of niobium. An 80\,nm thick layer of niobium is sputtered on a high resistivity silicon substrate, after HF cleaning and annealing at 700 degrees. The niobium is etched through a UV5 resist mask defined with electron-beam lithography. We use reactive-ion etching with an  NF$_3$ plasma, followed by an oxygen plasma ashing and subsequent chemical resist removal.
The SQUID is placed in a gap etched in the resonator center conductor.
We use e-beam lithography, double-angle evaporation of aluminum (50 and 70\,nm thick), and lift-off to fabricate the Josephson junctions.
In order to get a good contact between the niobium and the aluminum SQUID, an argon-ion milling step is realized in situ in the deposition chamber, prior to the evaporation of the two aluminum layers.

The sample is placed at the cold stage of a dilution refrigerator, at a temperature below 25\,mK, inside a magnetic shield. The magnetic flux in the SQUID loop is controlled both by a coil located on the sample box and with an on-chip current line. This line can be seen in Fig.~\ref{fig1}e. Whereas the coil is used for static tuning, the on-chip line has a large bandwidth, up to 12\,GHz, and thus allows fast control of the coupling. The microwave signal reflected from the sample undergoes heterodyne demodulation, and both  quadratures are digitized and sampled at 200\,MS/s.

\begin{table}[t]
  \begin{tabular}{c c c c c c c c c c}
 	\hline
     Sample & $f_1$ & $I_c$ & $C_\mathrm{out}$ & $C_\mathrm{c}$ & $\kappa$ & $g$ &  $Q_\mathrm{int}$  & $\tau_r$ & $\tau_s$ \\ \hline
      & GHz & $\mu$A & fF & fF & MHz & MHz &   & ns & $\mu$s\\ \hline
     A & 5.186 & 1.0 & 70 & 5.1 & 250 & 18.3 & 600\,000  & 14  & 18.4 \\ \hline
      B & 5.416 & 1.5 & 9.7 & 5.1 & 5 & 21.2 & 80\,000 & 200 & 2.4 \\ \hline
  \end{tabular} 
  \caption{Properties of the two measured devices: frequency of the storage resonator, critical current of the SQUID, capacitance between the coupling cavity and the transmission line, capacitance between the two resonators, coupling rate of the coupling cavity to the transmission line, coupling of the two resonators, internal quality factor of the storage cavity, minimum coupling time, and maximum storage time.}
  \label{table1}
\end{table}

First the devices are characterized with reflection measurements with a network analyzer as a function of the magnetic flux in the SQUID loop. The two coupled resonators give rise to two resonance modes, each of which lead to a Lorentzian-shaped signature in the reflection coefficient. From their linewidth, the coupling of these two modes to the transmission line can be extracted.

\begin{figure}[t]
\begin{center}
\includegraphics[width=1\columnwidth]{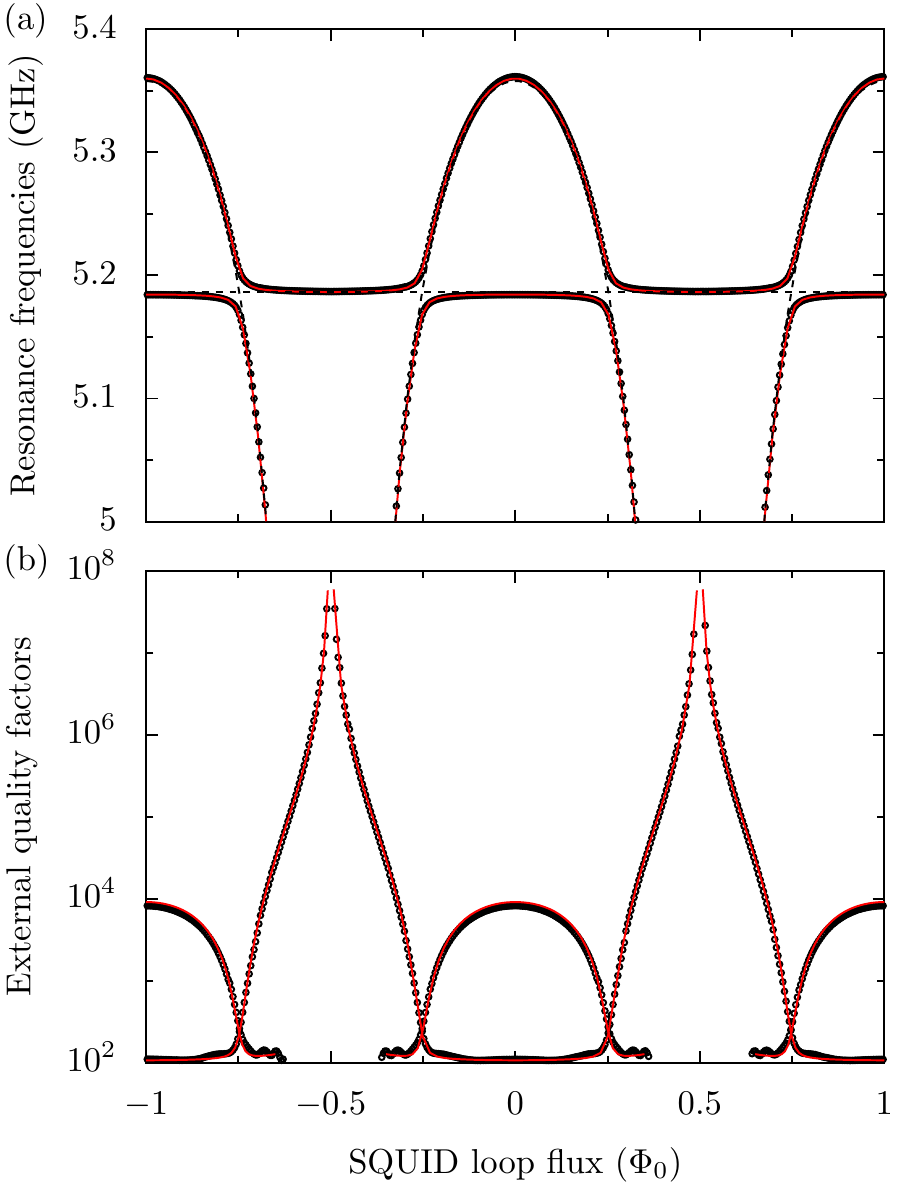}
\caption{Spectroscopy measurements of sample A.
a) Evolution of the frequencies of the two resonance modes as a function of flux in the SQUID loop. Plain, red lines: fit with an analytical model taking into account the coupling between the two resonators (see text). From the model, the resonance frequencies of the two cavities in absence of coupling can be computed (dashed lines).
b) External quality factors for both resonance modes, extracted from the measurements. The upper line varies over several order of magnitude, indicating the tunability of the coupling of the storage cavity with the detuning. The red line is the model explained in the text.} 
\label{fig2}
\end{center}
\end{figure}

Two different samples were tested. Their properties can be found in Table~\ref{table1}.
We present data on sample A, which was optimized for obtaining a large tunability range.
Sample B had a less favorable internal quality factor, probably due to small variations in the fabrication process. 
The natural frequency of the storage cavity is $f_{1A}=5.186$\,GHz. The frequency of the coupling cavity is periodically tuned with the flux (Fig.~\ref{fig2}a). It evolves from a maximum value of 5.36\,GHz obtained when the flux in the SQUID loop is an integer number of flux quanta. It reaches a minimum value for half integer flux quanta in the SQUID loop, well below 4\,GHz, the lower bound of our measurement band. The two cavities are on resonance at around a quarter of a flux quantum. This shows up as avoided level crossings, indicating that two modes arise from the coupling of the two cavities.
Fig.~\ref{fig2}a shows the extracted resonance frequencies $f_+$ and $f_-$ corresponding to these two modes. 
Their evolution is well reproduced with a simple analytical model
\begin{equation}
 f_\pm = \frac{1}{2}(f_1+f_2) \pm \sqrt{ g^2 + \left(\frac{\Delta f}{2}\right)^2  }
\end{equation}
where $f_2$ is the frequency of the coupling cavity, $\Delta f = f_2-f_1$ is the detuning, and $g = C_c Z_0 (2f_1^2 + f_2^2) = 18.3\,$MHz is the coupling between the two resonators. The frequency of the coupling cavity is given \cite{wallquist2006,sandberg2008} by $f_2  = f_2^0/(1+\gamma_0/\cos(\pi \Phi/\Phi_0))$,
where $f_2^0 = 5.810\,$GHz is the geometrical frequency of the coupling cavity, \textit{i.e.} as it would be without the SQUID. The participation ratio $\gamma_0 = 8.4\,$\% is the ratio between the Josephson inductance of the SQUID (at zero flux) and the inductance of the cavity.

Fig.~\ref{fig2}b shows the external quality factors extracted from the reflection measurement. They are defined as $Q_\mathrm{ext} = \omega_r /(2\Gamma_\mathrm{ext})$, where $\omega_r$ is the angular frequency of the resonance mode and $\Gamma_\mathrm{ext}$ the coupling rate to the transmission line. 
Far from the resonance points, the two lines can be interpreted separately in terms of coupling of each resonator to the transmission line. The bottom line with $Q_\mathrm{ext}\simeq 100$ is the result of the large, capacitive coupling of the coupling cavity, which is fixed and set by the value of the capacitance $C_\mathrm{out}$. On the contrary, the upper line, corresponding to the coupling of the storage cavity, strongly depends on the detuning between the two resonators. The coupling indeed varies over several orders of magnitude. The storage cavity therefore evolves from a strongly overcoupled to a strongly undercoupled regime as its external quality factor can be either larger or smaller than its intrinsic quality factor of $600\,000$. Close to zero detuning, the two modes tend to have equal coupling to the transmission line, which explains why the two lines cross at these points.

The evolution of the couplings with flux can be modeled with three free parameters, the two capacitances $C_\mathrm{out}$ and $C_\mathrm{c}$ and the SQUID inductance $L_\mathrm{S}$.
The transition rate describing the leakage of the energy contained in each mode is the average between the transition rate for each cavity, with weights taking into account how the coupled modes of the system decompose in the uncoupled cavity modes. This translates into the following equations:
\begin{align}
  \begin{aligned}
  \Gamma_\mathrm{ext}^+ &= \cos^2 \left( \frac{\theta}{2}\right) \Gamma_\mathrm{ext,1} + \sin^2 \left( \frac{\theta}{2}\right) \Gamma_\mathrm{ext,2} \\
  \Gamma_\mathrm{ext}^- &= \sin^2 \left( \frac{\theta}{2}\right) \Gamma_\mathrm{ext,1} + \cos^2 \left( \frac{\theta}{2}\right) \Gamma_\mathrm{ext,2}
  \end{aligned}
\end{align}
where $\theta=\arctan\left(2g/\Delta f \right)$ is the mixing angle describing the eigenmodes of the system. The model relies on analytical expressions for the coupling of the two cavities.
The coupling cavity has a coupling rate $\Gamma_\mathrm{ext,2}=\kappa/2=\omega_2(Z_0 \omega_2 C_\mathrm{out})^2/\pi$. The storage cavity has a residual coupling $\Gamma_\mathrm{ext,1}=(2/\pi)\omega_1(Z_0 \omega_1 C_\mathrm{c})^2( Z_0 \omega_1 C_\mathrm{out})^2/(1+ \left(\omega_1L_\mathrm{S}/Z_0\right)^2)$.
The best fit gives  $C_\textrm{c} = 5.1$\,fF, $C_\textrm{out} = 70$\,fF, and $L_\mathrm{S} = 345$\,pH.
At maximum detuning, the external quality factor of the storage cavity diverges because the SQUID not only detunes the coupling cavity but also enhances the scattering of the photons leaking out.

\begin{figure}[!t]
\begin{center}
\includegraphics[width=\columnwidth]{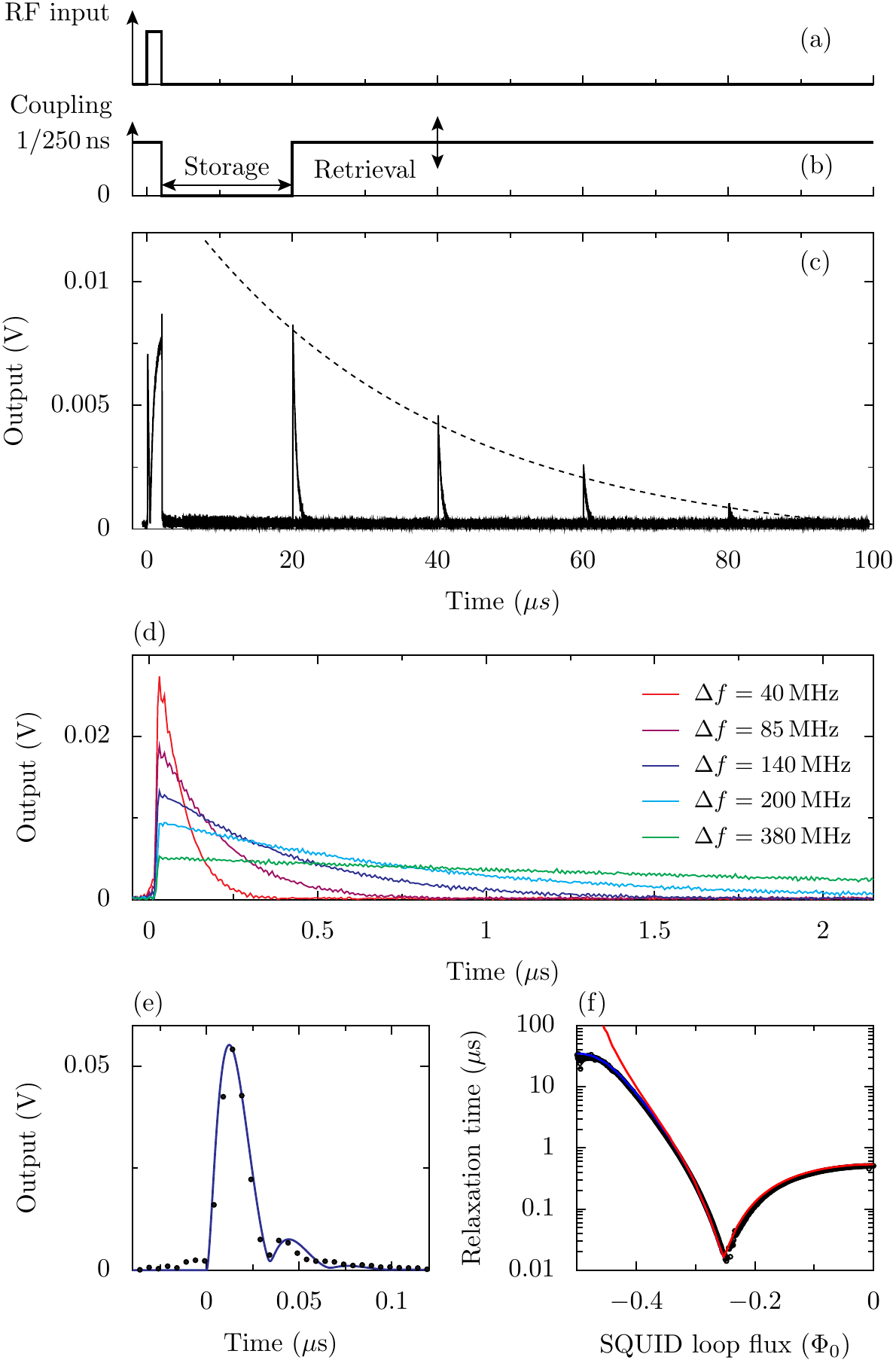}
\caption{Storage and release of microwaves with sample A. a) and b) Sketch of the control pulse sequence. A 2\,$\mu$s long microwave pulse on resonance with the storage cavity is sent at the input port. The coupling, initially at $1/250$\,ns, is set to zero for a variable storage time. Finally, the coupling is set back to its initial value, allowing the release of the stored energy.  c) Magnitude of the demodulated output signal for various storage times, ranging from 18 to 78\,$\mu$s. The input power is $-120$\,dBm, corresponding to storage of 80 photons on average. 
The recovered output amplitudes decrease with a characteristic amplitude decay time of 37\,$\mu$s (dashed line). d) Output traces obtained for several final detunings, associated with different couplings, or equivalently, different decay rates. 
e) Output trace obtained at zero detuning (black dots) showing fast but non exponential decay, and simulated trace (blue line) using numerical resolution of a Lindblad master equation with an initial coherent state in the storage cavity. The trace has been low-pass filtered with a cut-off frequency of 90\,MHz.
f) Output signal decay times extracted as a function of the final flux in the SQUID loop (black dots). The red line shows the same decay rate calculated from the coupling measurements shown in Fig.~\ref{fig2} (no fitting parameter). They differ only when the decay rate saturates because of the intrinsic loss rate of the storage cavity (blue line).}
\label{fig3}
\end{center}
\end{figure}

The static characterization of the system proves that the coupling of the storage cavity can be varied over several orders of magnitude. In particular this cavity can be efficiently decoupled from the transmission line, for a sufficient detuning  between the resonators, which enables to store a field inside. 
To demonstrate this, we performed time-resolved measurements, in which a field is built-up from an RF input pulse, stored, and released after a controlled delay.
Figs.~\ref{fig3}a and \ref{fig3}b show the experimental protocol.
Starting at an intermediate detuning, corresponding to a coupling of $1/250$\,ns, a microwave pulse on resonance with the storage cavity is applied to the input port of the device.
Once a steady-state field has been reached in the cavity, the detuning is increased to its maximum, with half a quantum of flux in the SQUID loop, and the input is turned off at the same time. The measured output signal goes to zero (Fig.~\ref{fig3}c), which proves that the coupling is effectively off. After a delay, the coupling is brought back to its initial value. The stored energy is therefore emitted to the transmission line, which is seen with a fast increase of the output signal, followed by an exponential decay as the field leaks out to the transmission line at a constant rate.

This experiment shows that the coupling can be turned on and off with our device. Moreover, it provides a measurement of the intrinsic lifetime of the cavity. Indeed, the longer the delay is, the less signal is recovered. Fitting the exponential decrease of the recovered signals, we estimate a lifetime of $18.4\,\mu$s or, equivalently, a quality factor of $6\times10^{5}$.
Note that this experiment has involved the storage of an average of 80 photons in the cavity. Moreover, we have checked that it can also be done at the single photon level, at the price of decreased signal to noise ratio.

To demonstrate that the coupling can be continuously adjusted to a desired value, the experimental protocol can be slightly modified. We now vary the detuning in the release step. As a result, we observe that the output signal decays exponentially at different rates, from a very slow decay when the detuning is large, to a very fast decay when the detuning is small (Fig.~\ref{fig3}d). The amplitude of the output signal also varies accordingly, since the same amount of energy is released in all experiments.
Fig.~\ref{fig3}f shows the decay time as a function of the flux in the SQUID loop in the release step. 
This is a direct measurement of the cavity lifetime, provided that a factor $1/2$ is taken into account in order to obtain the energy decay time from the voltage decay time.
At around half a quantum of flux, this time is large and saturates at the intrinsic lifetime of the cavity, which means that the energy is lost rather than being released to the transmission line.
At around a quarter of flux quantum, when the detuning approaches zero, the output signal no longer shows an exponential decay (Fig.~\ref{fig3}e), which prevents from relating the decay time to the lifetime of the cavity. When the two cavities are close to resonance, the stored microwave field oscillates between the two cavities. The release time is therefore limited by the transfer time, given by $1/4g=13.7$\,ns. 
Moreover, this oscillation is slightly underdamped, since the lifetime of the coupling cavity $1/\kappa$ is such that $\kappa/2\pi<4g$. It therefore shows up as an oscillation superimposed on the output voltage decay.  
This is well reproduced by a model of the release mechanisms, which includes low-pass filtering of the output signal to account for the finite bandwidth (90\,MHz) of the digitizer (Fig.~\ref{fig3}e).
Nevertheless, the experiment shows that the storage cavity lifetime varies over three orders of magnitude, from 14\,ns to 18\,$\mu$s.
Increasing $C_\textrm{out}$ such that $\kappa/2\pi$ reaches $4g$ would give a critically damped device which would not show any oscillation, and the release time would be limited by $4/\kappa$.

The tunable coupling system presented here has several advantages. It is easy to implement in any circuit-QED layout, where it can simply replace the usual coupling capacitances.
An extremely small coupling is naturally obtained when the coupling cavity is maximally detuned and the losses are limited by internal losses in the storage cavity. 
On the other hand, the maximum coupling is set by the two capacitances of the system.
Couplings even stronger than shown in this article should be easy to obtain.
Furthermore, with this coupling mechanism, the storage cavity is free from Josephson junctions, which cause additional dissipation in tunable resonators. Indeed, when the coupling is set to a small value in order to store a quantum field or perform coherent manipulation, no field exists in the tunable cavity.
This may explain why we reached a higher cavity lifetime than previously reported for a cavity equipped with a different, inductive tunable coupling \cite{yin2013}.

To conclude, we report on a superconducting circuit for coherent manipulation of microwave signals. 
Its primary function is to make the coupling between a microwave coplanar resonator and a transmission line tunable. This is interesting in circuit-QED experiments, for instance, to make single-photon on-demand sources or to shape photons. We demonstrated that the lifetime of a resonator can be tuned in a large range, from 14\,ns to 18\,$\mu$s. This proves that, while our system enables to achieve  large couplings, it does not enhance the cavity losses, and therefore allows to store microwaves for a long time.

We acknowledge the support of our colleagues at the Quantum Device Physics Laboratory and the Applied Quantum Physics Laboratory at Chalmers
University of Technology. 
We also acknowledge fruitful discussions with Christopher Wilson and financial support from the European Research Council, the European project PROMISCE and the Wallenberg Foundation.

\end{document}